\RequirePackage{fix-cm}
\documentclass[smallextended]{svjour3}       
\smartqed  
\usepackage{graphicx}
\usepackage{amsmath}
\usepackage{amssymb}
\usepackage{graphicx}
\usepackage{color}

\usepackage[compress]{natbib}
\bibpunct{[}{]}{,}{n}{,}{,} 

\usepackage[mathscr,scaled=1.15]{urwchancal}
\DeclareFontFamily{OT1}{pzc}{}
\DeclareFontShape{OT1}{pzc}{m}{it}%
{<-> s * [1.15] pzcmi7t}{}
\DeclareMathAlphabet{\mathpzc}{OT1}{pzc}{m}{it}

\journalname{Few-Body Systems}

\begin{document}

\title{The S- and P-wave low-lying baryons in the chiral quark model
%
}


\author{Gang Yang \and Jialun Ping \and Jorge Segovia}


\institute{Gang Yang \at
           Department of Physics and Jiangsu Key Laboratory for Numerical Simulation of Large Scale Complex Systems, Nanjing Normal University, Nanjing 210023, P. R. China \\
           \email{ygz0788a@sina.com}
\and
		   Jialun Ping \at
           Department of Physics and Jiangsu Key Laboratory for Numerical Simulation of Large Scale Complex Systems, Nanjing Normal University, Nanjing 210023, P. R. China \\
           \email{jlping@njnu.edu.cn}
\and
           Jorge Segovia \at
           IFAE and BIST, Universitat Aut\`onoma de Barcelona, E-08193 Bellaterra (Barcelona), Spain \\           
           \email{jsegovia@ifae.es}           
}

\date{\vspace*{-0.35cm} Received: date / Accepted: date}

\maketitle

\begin{abstract}
The $1S$, $2S$, $1P$ and $2P$ states of light baryons are investigated within the chiral quark model, paying particular attention to the well-known order reverse problem of $1P$ and $2S$ states. Besides a nonperturbative linear-screened confining interaction and a perturbative one-gluon exchange between quarks, we incorporate the Goldstone-boson exchanges taking into account not only the full octet of pseudoscalar mesons but also the scalar one. The scalar meson exchange potential simulates the higher order multi-pion exchange terms that appear in the chiral Lagrangian and its omission has been already admitted as a deficiency of the original model in describing, for instance, the $\rho-\omega$ splitting. The numerical approach to the three-body bound state problem is the so-called Gau\ss ian expansion method, which is able to get a precision as good as Faddeev calculations. With a set of parameters fixed to different hadron and hadron-hadron observables, we find that the chiral potential could play an important role towards the issue on the mass order reverse problem. We extend our calculation to the $qqQ$ and $qQQ$ sectors (with $q$ representing a light $u$-, $d$-, or $s$-quark and $Q$ denoting the charm quark or the bottom one) in which many new states have been recently observed. Some tentative assignments are done attending to the agreement between theoretical and experimental masses; however, we admit that other sources of information are needed in order to make strong claims about the nature of these states.
%
%
\end{abstract}


\newpage

\section{Introduction}

In 1963, L. D. Roper performed a partial-wave analysis in the process of pion-nucleon scattering and found a $P_{11}$ resonance~\cite{Roper:1964zza} whose Breit-Winger mass and width are set nowadays, respectively, to 1.43 GeV and 0.35 GeV~\cite{Patrignani:2016xqp}. This nucleon resonance was unexpected in naive quark models which incorporate an harmonic oscillator potential as the dominant interaction between quarks. This is because the zero-order energy spectrum is given by $E_n=\hbar\omega(N+3/2)$ with $N=2n_r+l$. Therefore, the nucleon's ground state ($J^P=1/2^+$) has $N=n_r=l=0$, the first negative parity state ($J^P=1/2^-$) appears with $n_r=0$ and $N=l=1$, and the next excited states have $N=2$ with either $n_r=1$, $l=0$ or $n_r=0$, $l=2$, and thus they have positive parity. Obviously, this is in contrast with the fact that $N(1535)$ $(J^P=1/2^-)$ is experimentally heavier than $N(1440)$ $(J^P=1/2^+)$.

During the past half century, a large amount of works, using different techniques, have tried to solve the so-called level ordering problem in the nucleon spectrum, i.e. to invert the mass ordering of the $N(1440)$ with respect the $N(1535)$. In the early time, Capstick and Isgur studied the baryon spectra with a relativized quark model in which confinement and one-gluon exchange (OGE) interaction were included~\cite{Capstick:1986bm}. Despite good description of baryon ground states, the issue with the order of energy levels for radial and orbital excited nucleons was unresolved, unless one added different ad-hoc values of mass shift for $1/2^-$ and $1/2^+$ $N^{\ast}$ states. Later on, Glozman and collaborators pointed out that, in the low energy regime of Quantum Chromodynamics (QCD), dynamical chiral symmetry breaking dictates the existence of Goldstone-boson exchange (GBE) interactions between the constituent quarks inside a hadron and thus they play an important role in the description of light mesons and baryons. Relying on the potentials of confinement and Goldstone-boson exchanges the spectrum of baryons was in good agreement with experimental data, especially for the mass order of nucleon resonances~\cite{Glozman:1997ag,Glozman:1997fs,Glozman:1999vd,Glozman:1995fu}. With a similar kind of spin-flavour interactions but working within the quark-diquark picture of a baryon, the authors of Ref.~\cite{Ferretti:2011zz} (see also~\cite{Santopinto:2014opa}) were able to reproduce the ordering of the $N(1440)$ and $N(1535)$ baryons, indicating that it is the nature of the interaction between quarks and not how they are clustered that matters when reproducing these masses. Further studies performed by Garcilazo {\it et al.}~\cite{Garcilazo:2001md,Garcilazo:2001ck,Garcilazo:2003wq}, using the Faddeev approach in momentum space to the three-body problem, showed that the relative position of the positive- and negative-parity states can be fixed by the interplay of relativistic kinematics and the one-pion-exchange interaction, playing also some role the perturbative one-gluon exchange potential.

The three valence-quark bound-state problem in continuum quantum field theory has been recently addressed within the Dyson-Schwinger equations (DSEs) formalism. References~\cite{Roberts:2011ym, Eichmann:2016yit, Eichmann:2016hgl, Lu:2017cln, Chen:2017pse} reveal that the $N(1440)$ is the first radial excitation of the nucleon whose unexpectedly low mass arises from a dressed-quark core that is shield by the so-called meson-cloud. In fact, Ref.~\cite{Segovia:2015hra} performed an analysis of the $\gamma^{(\ast)}N(940)\to N(1440)$ reaction and observed consistent results with the experimental data in the high $Q^2$-region where the the meson-baryon final-state interactions, i.e. meson-cloud components, are assumed to be highly suppressed. Similar conclusions have been obtained in Refs.~\cite{Cano:1998wz, Ramalho:2010js, Aznauryan:2007ja, Aznauryan:2012ec} analyzing the same reaction using quark models with greater or lesser level of sophistication, and in Refs.~\cite{Suzuki:2009nj, Kamano:2013iva} where meson-baryon final-state interactions are re-summed in dynamical coupled-channels models in order to transform a bare-baryon into the observed state. A review on the experimental and theoretical current status of the Roper resonance attending mostly on its electro-production transition form factors has been recently released~\cite{Burkert:2017djo}.

Lattice-regularized QCD studies of the Roper resonance have appeared lately~\cite{Wu:2016ixr, Lang:2016hnn} pointing out that meson-baryon components, specially $\pi\pi N$, in the Roper wave function play a more important role in its description than the naive 3-quark Fock component, with some results indicating that the $N(1440)$ could be a dynamically generated resonance~\cite{Liu:2016uzk, Wu:2016ixr}. However, it is worth to note: (i) many lattice calculations~\cite{Alexandrou:2014mka, Edwards:2011jj, Engel:2013ig, Liu:2014jua, Mahbub:2010rm} which report different results on the Roper issue have been ignored by the recent works, (ii) the 3-to-3 L\"uscher formalism is not yet firmly established (see review~\cite{Briceno:2017max} and references therein for an up-to-date status), and (iii) the pattern of chiral symmetry breaking should be implemented carefully on a lattice for the Roper problem~\cite{Liu:2016rwa}.

Acknowledging that continuum components on the Roper's wave function are important, coupled-channels calculations within the constituent quark model formalism have been performed obtaining results that sometimes are incompatible with each other. For example, Ref.~\cite{Obukhovsky:2011sc} finds the need of incorporating a large $\pi\pi N \equiv \sigma N$-component when studying the electro-production of the Roper resonance; whereas Ref.~\cite{JuliaDiaz:2006av} finds that the $qqq(q\bar q)$ component in the Roper ranges from $3\%$ to $25\%$ depending on the constituent quark mass while the $qqq(q\bar{q})^2$ components are negligible. A common feature of this kind of quark model calculations is that they demand a template of bare $qqq$ states in which one can trust in order to address safely the issues risen by coupling the meson-baryon continuum.

We present herein masses of the $1S$, $2S$, $1P$ and $2P$ states of light, charmed, doubly-charmed, single-bottom and double-bottom baryons, paying particular attention to the $1P$-$2S$ splittings in order to guess possible solutions of the so-called level ordering problem in the nucleon spectrum. Our chiral quark model (ChQM) contains Goldstone-boson exchange potentials, the perturbative one-gluon interaction and a linear-screened confining potential. In the meson-exchange potentials, we consider the full octet of pseudoscalar and scalar mesons. The later ones simulate the multi-pion exchange terms that appear in the chiral Lagrangian. They were incorporated in the original model~\cite{Vijande:2004he}\footnote{The interested reader is referred to Refs.~\cite{Valcarce:2005em, Segovia:2013wma} for detailed reviews on the naive quark model in which this work is based} for describing, e.g., the $\rho-\omega$ splitting~\cite{Vijande:2009pu} but not yet to address the level ordering problem~\cite{Garcilazo:2001md, Garcilazo:2001ck, Garcilazo:2003wq}. The three-body bound state problem is solved by means of the Gau\ss ian expansion method (GEM)~\cite{Hiyama:2003cu} which has been demonstrated to be as accurate as a Faddeev calculation (see Figs.~15 and~16 of Ref.~\cite{Hiyama:2003cu}). As it is well know, the quark model parameters are crucial in order to describe particular physical observables. We have used values that have been fitted before through hadron~\cite{Valcarce:1995dm, Vijande:2004he, Segovia:2008zza, Segovia:2008zz}, hadron-hadron ~\cite{Fernandez:1993hx, Valcarce:1994nr, Ortega:2009hj} and multiquark~\cite{Vijande:2006jf, Yang:2015bmv, Yang:2017rpg} phenomenology. Moreover, we have provided results with five different sets of model parameters in order to get some insight about the uncertainties associated with the model.

The observation of many new states in different heavy baryon sectors is another reason because we have focused our attention on the $qqQ$ and $qQQ$ spectra, with $q$ representing a light $u$-, $d$-, or $s$-quark and $Q$ denoting either $c$- or $b$-quark. Some tentative assignments are done attending to the agreement between theoretical and experimental masses. However, we admit that other sources of information, such as total widths and decay patterns, are needed in order to make strong claims about the nature of the states.

The structure of the paper is as follows. In Sec.~\ref{sec:model} the ChQM, baryon wave-functions and GEM are briefly presented and discussed. Section~\ref{sec:result} is devoted to presenting our results. We finish summarizing and giving some conclusions in Sec.~\ref{sec:summary}.


\section{Theoretical framework}
\label{sec:model}

The ChQM is based on the fact that a nearly massless current light quark acquires a dynamical, momentum-dependent mass, namely, the constituent quark mass due to its interaction with the gluon medium. To preserve chiral invariance of the QCD Lagrangian new interaction terms, given by Goldstone-boson exchanges, should appear between constituent quarks. Therefore, the chiral part of the quark-quark interaction can be expressed as follows
\begin{equation}
V_{\text{OBE}}(\vec{r}_{ij}) =  V_{\pi}(\vec{r}_{ij}) + V_{K}(\vec{r}_{ij}) + V_{\eta}(\vec{r}_{ij}) +
V_{\text{sc}}(\vec{r}_{ij}).
\end{equation}
The different terms of the OBE potential contain central and tensor or central and spin-orbit contributions; only the central ones will be considered attending the goal of the present manuscript and for clarity in our discussion. Detailed expressions for $V_{\pi}$, $V_{K}$ and $V_{\eta}$ can be found, for instance, in Ref.~\cite{Yang:2017rpg}. The scalar potential, $V_{\text{sc}}(\vec{r}_{ij})$, considers not only the leading contribution of the 2-pion exchange interaction in the isoscalar-scalar channel but also other higher multi-pion terms that are simulated through the exchange between two constituent quarks of the full octet of scalar mesons:
\begin{align}
V_{\text{sc}}(\vec{r}_{ij}) &= V_{a_{0}}(\vec{r}_{ij}) \sum_{a=1}^{3} \lambda_{i}^{a} \cdot \lambda_{j}^{a} +
V_{\kappa}(\vec{r}_{ij}) \sum_{a=4}^{7} \lambda_{i}^{a}\cdot\lambda_{j}^{a} \nonumber \\
&
+ V_{f_{0}}(\vec{r}_{ij}) \lambda_{i}^{8}\cdot\lambda_{j}^{8} + V_{\sigma}(\vec{r}_{ij}) \lambda_{i}^{0} \cdot
\lambda_{j}^{0} \,,
\end{align}
where the radial form is the same for all of them~\cite{Yang:2017rpg}, but there is a different $SU(3)$-flavor operational dependence: the $\lambda^a$ with $a=1,\ldots, 8$ are the $SU(3)$-flavor Gell-Mann matrices and $\lambda^0$ is just the $3\times3$ identity matrix multiplied by a factor of $\sqrt{2/3}$ which is according to the property of Gell-Mann matrices.

Confinement is one of the crucial aspects of the strong interaction that is widely accepted and incorporated into any QCD based model. It is believed that multigluon exchanges produce an attractive linearly rising potential proportional to the distance between quarks. This idea has been confirmed, but not rigorously proved, by quenched lattice gauge calculations applied to infinitely heavy valence quark systems~\cite{Bali:2000gf}. However, sea quarks are also an important ingredient of the strong interaction dynamics. When they are included in the lattice calculations they contribute to the screening of the rising potential at low momenta and eventually to the breaking of the binding string~\cite{Bali:2005fu}. These features have been taken into account in our model including the following expression for the confinement potential
\begin{equation}
V_{\rm CON}(\vec{r}_{ij})=\left[-a_{c}(1-e^{-\mu_{c}r_{ij}})+\Delta \right]
(\vec{\lambda}_{i}^{c}\cdot\vec{\lambda}_{j}^{c}) \,,
\end{equation}
where $a_{c}$, $\mu_{c}$ and $\Delta$ are parameters; $\Delta$ is a global constant fixing the origin of energies and $\vec{\lambda}^{c}$ are the $SU(3)$-color matrices. At short distances this potential presents a linear behaviour with an effective confinement strength $\sigma = -a_{c}\,\mu_{c}\,(\vec{\lambda}^{c}_{i}\cdot \vec{\lambda}^{c}_{j})$, while it becomes constant at large distances. Note that the Lorentz character of the confinement has not yet firmly established, it determines the associated spin-dependent terms of the interaction. In our case, we are considering just the central term and thus this issue is avoided.

Beyond the nonperturbative energy scale, $\Lambda_{\text{QCD}}$, one expects that the dynamics of the bound-state system is governed by QCD perturbative effects. We take it into account through a standard color Fermi-Breit interaction called one-gluon exchange that is obtained from the following vertex Lagrangian
\begin{equation}
\mathcal{L}_{\rm qqg}=i\sqrt{4\pi\alpha_{s}}\,\bar{\psi}\gamma_{\mu}G^{\mu}_{c}
\lambda^{c}\psi \,,
\end{equation}
with $G^\mu_c$ the gluon field and $\alpha_{s}$ the strong coupling constant. The central potential derived from the Lagrangian is given by
\begin{equation}
V_{\rm OGE}(\vec{r}_{ij}) = \frac{1}{4} \alpha_{s} (\vec{\lambda}_{i}^{c}\cdot \vec{\lambda}_{j}^{c}) \Big[
\frac{1}{r_{ij}} -\frac{(\vec{\sigma}_{i}\cdot\vec{\sigma}_{j})}{6m_{i}m_{j}}
\frac{e^{-r_{ij}/r_{0}(\mu)}}{r_{ij}r_{0}^{2}(\mu)} \Big] \,,
\end{equation}
where the contact term has been regularized as
\begin{equation}
\delta(\vec{r}_{ij})\sim\frac{1}{4\pi r_{0}^{2}}\frac{e^{-r_{ij}/r_{0}}}{r_{ij}} \,.
\end{equation}

For the three-body bound-state system, the general form of the Hamiltonian is given by
\begin{equation}
H = \sum_{i=1}^{3}\left( m_i+\frac{\vec{p\,}^2_i}{2m_i}\right) - T_{\text{CM}} + \sum_{j>i=1}^{3} V(\vec{r}_{ij}) \,,
\label{Ham}
\end{equation}
where each quark is considered nonrelativistic, $T_{\text{CM}}$ is the center-of-mass kinetic energy and the two-body potential includes, as already mentioned, the central terms of confining, one-gluon and Goldstone-boson interactions.

The model parameters have been fixed in advance reproducing hadron~\cite{Valcarce:1995dm, Vijande:2004he, Segovia:2008zza, Segovia:2008zz}, hadron-hadron ~\cite{Fernandez:1993hx, Valcarce:1994nr, Ortega:2009hj, Ortega:2016mms} and multiquark~\cite{Vijande:2006jf, Yang:2015bmv, Yang:2017rpg} phenomenology. However, we choose five sets of parameters that are listed in Table~\ref{tab:Parameters} in order to get some insight about the uncertainties associated with the model. Particularly interesting is the use of sets II, IV and V, where an effective scale-dependent strong coupling constant: $\alpha_s(\mu_{ij})=\alpha_0/\ln[(\mu_{ij}^2+\mu_0^2)/\Lambda_0^2]$, is implemented. All terms in the former expression are model parameters except $\mu_{ij}$, which is the reduced mass of the quark--(anti-)quark pair affected by the interaction. This parametrization is useful in order to get a consistent description of light, strange and heavy hadrons~\cite{Vijande:2004he}.

\begin{table*}
\begin{center}
\caption{\label{tab:Parameters} ChQM with five sets of parameters.}
\scalebox{0.75}{\begin{tabular}{ccccccc}
\hline\noalign{\smallskip}
                 &                     & set I & set II & set III  & set IV  & set V \\
\noalign{\smallskip}\hline\noalign{\smallskip}
                 & $m_u$=$m_d$ (MeV)   &  313  &  313   &  313 &  313  &  313\\
Quark mass       & $m_s$ (MeV)         &  555  &  555   &  555 &  555  &  555\\
                 & $m_c$ (MeV)         & 1660  & 1620   & 1580 &1540  &1520\\
                 & $m_b$ (MeV)         & 5030  & 5030   & 4930  &4930  &4930\\
\hline
                 & $a_c$ (MeV)          & 253.1    & 202.1 & 461.3 &210.35  &180.71\\
Confinement      & $\mu_c$ (fm$^{-1})$  & 0.466  &   0.677 & 0.570 &0.35451  &0.90462\\
                 & $\Delta$ (MeV)       & 67.74  &  62.45  & 164.52 &11.634  &52.693\\
\hline
                 & $\alpha_s^{uu}~$     & 0.576  & $\alpha_0$=0.880                    & 0.477   & $\alpha_0$=0.71737   &$\alpha_0$=0.63736\\
                 & $\alpha_s^{us}~$     & 0.576  & $\Lambda_0$=1.8445 $\text{fm}^{-1}$ & 0.459   & $\Lambda_0$= 2.4686 $\text{fm}^{-1}$  &$\Lambda_0$= 2.583  $\text{fm}^{-1}$\\
                 & $\alpha_s^{ss}~$     & 0.576  & $\mu_0$=659.93 MeV              & 0.359  & $\mu_0$= 754.56 MeV  &$\mu_0$= 777.75 MeV\\
                 & $\alpha_s^{uc}~$     & 0.553  & -                                   & 0.221   & -  & -\\
   OGE       & $\alpha_s^{sc}~$     & 0.553  & -                                   & 0.210   & -  & - \\
                 & $\alpha_s^{cc}~$     & 0.542  & -                                   & 0.203   & -  & - \\
                 & $\alpha_s^{ub}~$     &0.53   & -                                   &0.138   & -  & -  \\
                 & $\alpha_s^{sb}~$     &0.53   & -                                   &0.113   & -  & -  \\
                 & $\alpha_s^{bb}~$     &0.48   & -                                   &0.091   & -  & -  \\
                 & $\hat{r}_0~$(MeV~fm) & 30.86 & 40.73                     & 37.19   &123.27  &168.98\\
\hline
                 & $m_{\pi}$ (fm$^{-1}$)                      & 0.70  & 0.70 & 0.70 & 0.70 & 0.70\\
                 & $m_{K}$ (fm$^{-1}$)                        & 2.51  & 2.51 & 2.51 &2.51 &2.51\\
                 & $m_{\eta}$ (fm$^{-1}$)                     & 2.77  & 2.77 & 2.77  &2.77  &2.77\\
                 & $\Lambda_\pi=\Lambda_{\sigma}$ (fm$^{-1}$) & 4.20  & 4.20 & 4.20  &5.20  &5.20\\
Goldstone boson  & $\Lambda_\eta$ (fm$^{-1}$)       & 5.20  & 5.20 & 5.20  &6.20  &6.20\\
                 & $\Lambda_K$ (fm$^{-1}$)       & 5.20  & 5.20 & 5.20  &7.20  &7.20\\
                 & $\theta_P(^\circ)$                         & -15   & -15  & -15 & -15 & -15\\
                 & $g^2_{ch}/(4\pi)$                          & 0.54  & 0.54 & 0.54  &0.7  &0.7\\
\hline
SU(3)                       & $m_{\sigma}$ (fm$^{-1}$)  & 3.42 & 3.42 & 3.42 &3.42&3.42\\
Scalar nonet                & $\Lambda_{s}$ (fm$^{-1}$) & 5.20 & 5.20 & 5.20 &6.20 &6.20\\
$s=\sigma, a_0,\kappa ,f_0$ & $m_{s}$ (fm$^{-1}$)       & 4.97 & 4.97 & 4.97 &4.97 &4.97\\
\noalign{\smallskip}\hline
\end{tabular}}
\end{center}
\end{table*}

As for the baryon's wave function, each quark has color, spin, flavor and spatial degrees-of-freedom. According to the empirical fact that color sources have never seen as isolated particles, the color wave function of a baryon can be easily written as
\begin{equation}
\chi^{c} = \frac{1}{\sqrt{6}} (rgb - rbg + gbr - grb + brg - bgr) \,.
\end{equation}
The spin wave function of a $3$-quark system taking into account any possible quantum number combination is as below,
\begin{align}
\chi_{\frac32,\frac32}^{\sigma}(3)  &= \alpha\alpha\alpha \,, \\
\chi_{\frac32,\frac12}^{\sigma}(3)  &= \frac{1}{\sqrt{3}}(\alpha\alpha\beta+\alpha\beta\alpha+\beta\alpha\alpha) \,,
\end{align}
\begin{align}
\chi_{\frac32,-\frac12}^{\sigma}(3) &= \frac{1}{\sqrt{3}} (\alpha\beta\beta+\beta\alpha\beta+\beta\beta\alpha) \,, \\
\chi_{\frac32,-\frac32}^{\sigma}(3)  &= \beta\beta\beta \,, \\
\chi_{\frac12,\frac12}^{\sigma 1}(3) &= \frac{1}{\sqrt{6}} (2\alpha\alpha\beta-\alpha\beta\alpha-\beta\alpha\alpha) \,, \\
\chi_{\frac12,\frac12}^{\sigma 2}(3) &= \frac{1}{\sqrt{2}} (\alpha\beta\alpha-\beta\alpha\alpha) \,, \\
\chi_{\frac12,-\frac12}^{\sigma 1}(3) &= \frac{1}{\sqrt{6}} (\alpha\beta\beta-\alpha\beta\beta-2\beta\beta\alpha) \,, \\
\chi_{\frac12,-\frac12}^{\sigma 2}(3) &= \frac{1}{\sqrt{2}} (\alpha\beta\beta-\beta\alpha\beta) \,.
\end{align}
The charm and bottom quarks are much heavier than the light ones: $u$, $d$ and $s$ quark. Therefore, we investigate the baryon with quark content $u$, $d$, $s$ and $c$ or $b$ in $SU(3)$-flavor case and the corresponding flavor wave functions are given by
\begin{align}
N^1 &= \frac{1}{\sqrt{6}}(2uud-udu-duu) \,, \\
N^2 &= \frac{1}{\sqrt{2}}(ud-du)u \,, ~~~~ \Delta = uuu \,, \\
\Lambda^1 &= \frac{1}{2}(usd-dsu+sud-sdu) \,, \\
\Lambda^2 &= \frac{1}{\sqrt{12}}(2uds-2dus+usd-dsu-sud+sdu) \,, \\
\Sigma^1 &= \frac{1}{\sqrt{12}}(2uds+2dus-usd-dsu-sud-sdu) \,, \\
\Sigma^2 &= \frac{1}{2}(usd-sud+dsu-sdu) \,, \\
\Sigma^* &= \frac{1}{\sqrt{6}}(uds+usd+dus+dsu+sud+sdu), \\
\Xi^1 &= \frac{1}{\sqrt{6}}(uss+sus-2ssu) \,, \\
\Xi^2 &= \frac{1}{\sqrt{2}}(us-su)s, \\
\Xi^* &= \frac{1}{\sqrt{3}}(uss+sus+ssu) \,, \\
\Omega &= sss \,,
\end{align}
for the light baryons and
\begin{align}
\Lambda_Q &= \frac{1}{2}(ud-du)Q \,, \\
\Sigma_Q &= \frac{1}{2}(ud+du)Q \,,
\end{align}
\begin{align}
\Xi_Q &= \frac{1}{2}(us-su)Q \,, \\
\Xi_Q' &= \frac{1}{2}(us+su)Q \,, \\
\Omega_Q &= ssQ \,, \\ 
\Xi_{QQ} &= uQQ \,, \\
\Omega_{QQ} &= sQQ \,,
\end{align}
for heavy-flavored baryons where $Q$ represents either $c$- or $b$-quark.

Among the different methods to solve the Schr\"odinger-like 3-body bound state equation we use the Rayleigh-Ritz variational principle, which is one of the most extended tools to solve eigenvalue problems due to its simplicity and flexibility. However, it is of great importance how to choose the basis on which to expand the wave function. The spatial wave function of a $3$-quark system is written as follows:
\begin{equation}
\psi_{LM_L} = \left[ \phi_{n_1l_1m_{l_1}}(\vec{r}\,) \phi_{n_2l_2m_{l_{2}}}(\vec{R}\,) \right]_{LM_L} \,.
\label{w3q}
\end{equation}
where the two Jacobi coordinates in Eq.~(\ref{w3q}) are defined as
\begin{equation}
\vec{r} = \vec{x}_1 - \vec{x}_2 \,, ~~~
\vec{R} = \vec{x}_3 - \frac{m_1 \vec{x}_1 + m_2 \vec{x}_2}{m_1+m_2} \,.
\end{equation}
This choice is convenient because, for a nonrelativistic system, the center-of-mass kinetic term $T_{\text{CM}}$ can be completely eliminated. In order to make the calculation tractable, even for complicated interactions, we replace the orbital wave functions, $\phi_{nlm}$, by a superposition of infinitesimally-shifted Gaussians (ISG)~\cite{Hiyama:2003cu}:
\begin{equation}
\begin{split}
\phi_{nlm}(\vec{r}\,) &= N_{nl} r^{l} e^{-\nu_{n} r^2} Y_{lm}(\hat{r}) \\
&
= N_{nl} \lim_{\varepsilon\to 0} \frac{1}{(\nu_{n}\varepsilon)^l} \sum_{k=1}^{k_{\rm
max}} C_{lm,k} e^{-\nu_{n}(\vec{r}-\varepsilon \vec{D}_{lm,k})^{2}} \,,
\end{split}
\end{equation}
where the limit $\varepsilon\to 0$ must be carried out after the matrix elements have been calculated analytically. This new set of basis functions makes the calculation of three- and, in general, few-body matrix elements very easy without the laborious Racah algebra. Moreover, all the advantages of using Gau\ss ians remain with the new basis
functions.

Following Ref.~\cite{Hiyama:2003cu}, we employ Gau\ss ian trial functions whose ranges are in geometric progression. This enables the optimization of the basis employing a small number of free parameters. Moreover, the geometric progression is dense at short distances so that it allows the description of the dynamics mediated by short range potentials. The fast damping of the Gau\ss ian tail is not a problem, since we can choose the maximal range much longer than the hadronic size.

We have constructed explicitly an antisymmetric wave function for only two particles of the 3-body system by choosing the appropriate symmetries of color, spin, flavor and spatial degrees of freedom. Just coupling the third particle to the other two particles with appropriate Clebsch-Gordan coefficients does not produce the totally anti-symmetric wave function. Therefore, we need to act the antisymmetric operator of the $3$-quark system, ${\cal A}$, on the combined color, spin-flavor, and spatial wave function. The complete antisymmetric wave function is written as
\begin{equation}
\Psi_{JM_{J}}={\cal A} \left[ \left[ \psi_{LM_L} \chi^{\sigma}_{SM_S}(3) \right]_{JM_J} \chi^{f} \chi^{c} \right] \,.
\label{wf}
\end{equation}
There should be six terms in ${\cal A}$ for a system with three identical particles but if one constructs an antisymmetric wave function for the first two quarks, the operator is simplified to just three terms:
\begin{equation}
{\cal A} = 1 - (13) - (23) \,.
\end{equation}


\begin{table}[!t]
\begin{center}
\caption{\label{tab:LightBaryons} Masses, in MeV, of $1S$, $2S$, $1P$ and $2P$ light baryons predicted by the ChQM and using sets I-V of model parameters. Experimental data are from Ref.~\cite{Patrignani:2016xqp}.}
\begin{tabular}{ccccccc}
\hline\noalign{\smallskip}
$N(939)$ & set I & set II & set III & set IV & set V & Exp. \\
$1S$ &  939 & 939 & 939 & 936 & 956  &939 \\
$2S$ & 1436 & 1426 & 1678 & 1461 & 1493  & 1440 \\
$1P$ & 1411 & 1417 & 1597 & 1481 & 1521  & 1535\\
$2P$ & 1670 & 1617 & 2043 & 1712 & 1666  & 1650 \\[1.5ex]
$\Lambda(1116)$ & set I & set II & set III & set IV & set V & Exp. \\
$1S$ & 1127 & 1128 & 1111 & 1103 & 1117 & 1116\\
$2S$ & 1618 & 1617 & 1829 & 1638 & 1675 & 1600\\
$1P$ & 1592 & 1604 & 1742 & 1661 & 1703 & 1670\\
$2P$ & 1853 & 1817 & 2161 & 1894 & 1871 & 1800\\[1.5ex]
$\Sigma(1193)$ & set I & set II & set III & set IV & set V & Exp. \\
$1S$ & 1263 & 1269 & 1254 & 1319 & 1333 & 1193\\
$2S$ & 1696 & 1694 & 1913 & 1741 & 1767 & 1660 \\
$1P$ & 1652 & 1664 & 1804 & 1737 & 1774 & 1580 \\
$2P$ & 1909 & 1908 & 2223 & 1971 & 1997 & 1750 \\[1.5ex]
$\Xi(1318)$ & set I & set II & set III & set IV & set V & Exp. \\
$1S$ & 1386 & 1395 & 1373 & 1391 & 1399 & 1318\\
$2S$ & 1837 & 1846 & 2028 & 1872 & 1906 &1950  \\
$1P$ & 1798 & 1819 & 1925 & 1885 & 1922 & 1820 \\
$2P$ & 2066 & 2082 & 2336 & 2123 & 2144 & 2030 \\[1.5ex]
$\Delta(1232)$ & set I & set II & set III & set IV & set V & Exp. \\
$1S$ & 1232 & 1236 & 1233 & 1222 & 1224 & 1232 \\
$2S$ & 1613 & 1589 & 1861 & 1609 & 1612 & 1600 \\
$1P$ & 1522 & 1523 & 1697 & 1561 & 1590 & 1620 \\
$2P$ & 1708 & 1699 & 2156 & 1810 & 1754 & 1900 \\[1.5ex]
$\Sigma^\ast(1385)$ & set I & set II & set III & set IV & set V & Exp. \\
$1S$  & 1391 & 1398 & 1362 & 1384 & 1376 & 1385 \\
$2S$  & 1778 & 1766 & 1988 & 1781 & 1792 & 1840 \\
$1P$  & 1652 & 1664 & 1804 & 1737 & 1774 & 1580 \\
$2P$  & 1909 & 1886 & 2223 & 1971 & 1997 & 1750 \\[1.5ex]
$\Xi^\ast(1530)$ & set I & set II & set III & set IV & set V & Exp. \\
$1S$ & 1536 & 1549 & 1493 & 1535 & 1517 & 1530 \\
$2S$ & 1930 & 1932 & 2111 & 1947 & 1961 & 1950 \\
$1P$ & 1798 & 1819 & 1925 & 1885 & 1922 & 1820\\
$2P$ & 2066 & 2082 & 2336 & 2123 & 2144 & 2030 \\[1.5ex]
$\Omega(1673)$ & set I & set II & set III & set IV & set V & Exp. \\
$1S$ & 1663 & 1687 & 1631 & 1685 & 1655  & 1673 \\
$2S$ & 2068 & 2087 & 2236 & 2108 & 2124 & -\\
$1P$ & 1977 & 2010 & 2073 & 2059 & 2080 & -\\
$2P$ & 2259 & 2202 & 2525 & 2316 & 2316 & -\\
\noalign{\smallskip}\hline
\end{tabular}
\end{center}
\end{table}

\section{Results and discussion}
\label{sec:result}

The main goal of this work is to investigate if the chiral quark model is able to locate the first radial excitation of the nucleon (positive parity state) below its ground state $P$-wave partner (negative parity state). This study has been performed before in Refs.~\cite{Garcilazo:2001md, Garcilazo:2001ck, Garcilazo:2003wq} within a very similar formalism but without taking into account in the Goldstone-boson exchange interaction the scalar mesons which belong to the same flavor-octet than the $\sigma$-meson. The inclusion of the one-boson exchange potentials associated with the full scalar octet mesons has been demonstrated to be determinant for improving the phenomenology of mesons~\cite{Vijande:2009pu} and even baryons~\cite{Garcilazo:2003wq}. Moreover, for simplicity, Refs.~\cite{Garcilazo:2001md, Garcilazo:2001ck, Garcilazo:2003wq} considered just a linear confining interaction whereas the linear-screened potential has been applied with great success to the description of highly excited light~\cite{Segovia:2008zza}, heavy-light~\cite{Segovia:2015dia, Ortega:2016mms, Ortega:2016pgg} and heavy~\cite{Segovia:2010zzb, Segovia:2011zza, Segovia:2016xqb, Ortega:2016hde, Yang:2015bmv} mesons, even including multiquark configurations. These are the main motivations to re-visit the so-called level ordering problem within this formalism.

We report in Table~\ref{tab:LightBaryons} the predicted masses for ground and radially excited states, with either $L=0$ or $L=1$ orbital angular momentum, of the octet and decuplet light baryons. We show our results using sets I-V of model parameters and compare them with experimental data if available. One can conclude the following:
\begin{itemize}
\item[(i)] The states are located higher in the spectrum when going from set I to set V. This is mostly due to larger values of the effective string tension. The change in mass is less than $100\,\text{MeV}$.
\item[(ii)] The ground states of the octet and decuplet light baryons are reasonably well described (see, for instance, the second column of Table~\ref{tab:LightBaryons}). The biggest discrepancy appears in the mismatch between theory and experiment for $\Sigma(1193)$ and $\Xi(1318)$. Their masses are predicted around $70\,{\text{MeV}}$ higher than experiment and it is due to the hyperfine interaction. We have not fine-tuned such interaction because our goal is to observe, in the most cleaning way, the location of the orbital excitation with respect the first radially excited state.
\item[(iii)] The first radial excitation of the nucleon is predicted above its orbitally excited one for sets I, II, and III. However, a right mass ordering with respect experimental data is obtained when using sets IV and V. In these sets, the parameters of Goldstone-boson exchange interactions are adjusted properly and our results go in line with those of Refs.~\cite{Garcilazo:2001md, Garcilazo:2001ck, Garcilazo:2003wq} which conclude that the pseudoscalar, confining and color Fermi-Breit interactions compete for mass-splittings.
\item[(iv)] The incorporation of the so-called $\sigma$-, $f_{0}$-, $a_{0}$- and $\kappa$-exchange potentials gives us some flexibility to get a better global description of the spectrum of the light baryons.
\end{itemize}

Attending to Table~\ref{tab:Parameters}, within our formalism, the first radial excitation of the nucleon is below its orbitally excited partner when the chiral coupling constant ($g_{ch}$) and the Goldstone-boson mass scales ($\Lambda_\chi$) are slightly augmented. Namely, in columns 5 and 6 of Table~\ref{tab:LightBaryons} one can see that the mass of the nucleon's $2S$ state ranges from $1.46\,\text{GeV}$ to $1.49,\text{GeV}$ whereas the mass of its first orbital excited state goes from $1.48\,\text{GeV}$ to $1.52\,\text{GeV}$.

In order to support the idea that chiral symmetry and its breaking pattern in QCD is responsible of inverting the $1P-2S$ mass splitting, we have performed a calculation of the $2S$ and $1P$ states using sets IV and V of model parameters and turning off the Goldstone-boson exchange interactions between quarks. Our results are 1698 MeV and 1654 MeV using set IV and 1698 MeV and 1621 MeV with set V. Therefore, the first radial excitation of the nucleon is located above its first orbitally excited state and thus our claim is that the so-called ordering problem was related with quark models that do not incorporate Goldstone-boson exchanges because the perturbative one-gluon exchange interaction was not enough to reverse the energy location of the $N=1$ and $N=2$ bands in the pure harmonic limit.

For the $\Lambda$-baryon, the calculated mass of its $2S$ state is always higher than the $1P$ state when using the first three sets of parameters. However, using the last two sets, we manage to locate the first radial excitation below the first orbitally excited one. The obtained mass of $2S$ state is within the interval $[1.64,1.68]\,\text{GeV}$, and the $1P$ state is predicted to have a mass between $1.66\,\text{GeV}$ and $1.70\,\text{GeV}$. Looking at, for example, column~5 of Table~\ref{tab:LightBaryons}, our preferred assignment for the $2S$ state is $\Lambda(1600)$ and for the $1P$ state is $\Lambda(1670)$. There are two extra states measured experimentally: $\Lambda(1405)$ and $\Lambda(1520)$. In our approach, these two states cannot be described as naive three-quark states and thus higher Fock components must be invoked. There is a global agreement among the scientific community that the $\Lambda(1405)$ is a dynamically generated resonance produced in the $NK$ scattering process~\cite{Oset:1997it, Jido:2003cb}. The $\Lambda(1520)$ can be described as a meson-baryon quasi-bound state in the chiral unitary model~\cite{Hyodo:2006uw}, but its nature is still under discussion with some proposed reactions, $\gamma^\ast N\to \Lambda(1520)K$ and $\gamma^\ast N\to \Lambda(1520)K^{\ast}$, able to discern among $3$-quark and meson-baryon possibilities.

As to the $\Sigma$-baryon in the octet multiplet, the theoretical masses for the $1P$ and $2S$ states using the first two sets of model parameters are about $1.65\,\text{GeV}$ and $1.69\,\text{GeV}$, respectively. Therefore, the masses compare nicely with the experimental ones associated with the $\Sigma(1620)$ and $\Sigma(1660)$ baryons. Note, however, that we can inverse the ordering of the $2S$ and $1P$ states when using the sets IV and V but the predicted masses are so close to each other that a definitive statement cannot be made. It is interesting to observe that using the set III of model parameters we obtain masses for the radial- and orbital-excited states which are much higher than the ones predicted by other groups; even higher than the experimental mass of the $\Sigma(1775)$ ($J^P=5/2^-$) indicating that the $\Sigma$ baryon can not be described well using set III of model parameters.

Higher masses than experimental data are found for the $\Xi$-baryon. However, the experimental situation is not yet clear; for example, the spin-parity quantum numbers of the excited $\Xi$ baryons above $1.82$ GeV are not fixed. We think that this issue can be solved in the near future by fine-tuning the slope of the confining potential at short distances and by modifying slightly the mass of the strange quark.

We have already mentioned that the ground states of the $\Delta$, $\Sigma^*$, $\Xi^*$ and $\Omega$ baryons are reasonably well described in our formalism. It is also worth to note that for all decuplet light baryons the corresponding first radial excitation is located above the first orbitally excited state. The $\Omega$-baryon deserves some special attention, there are four states collected in the PDG whose last measurements date from the 1980s. We predict a mass for the $2S$ state which goes from $2.07$ to $2.23$~GeV, and it appears as a natural candidate for the $\Omega(2250)$ baryon. The $1P$ state is located within the energy interval $1.98-2.08$~GeV, and thus we consider that it is missing experimentally. In any case, more theoretical and experimental work is needed in order to clarify the situation of $\Omega$ states.

\begin{figure}[!t]
\begin{center}
\includegraphics[clip,width=0.49\textwidth,height=0.30\textheight]{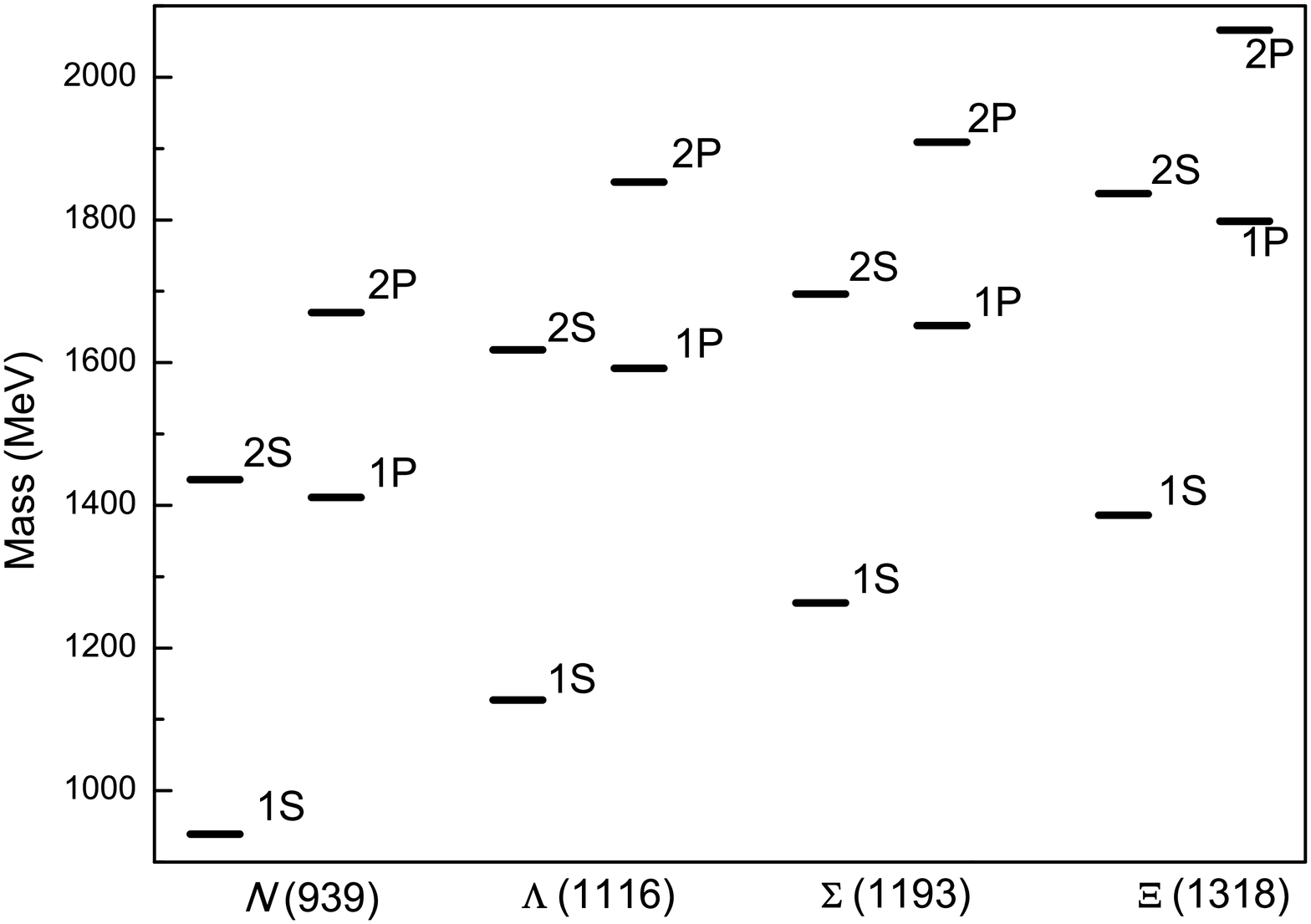}
\put(-130,130){(a)}
\put(35,130){(b)}
\includegraphics[clip,width=0.49\textwidth,height=0.30\textheight]{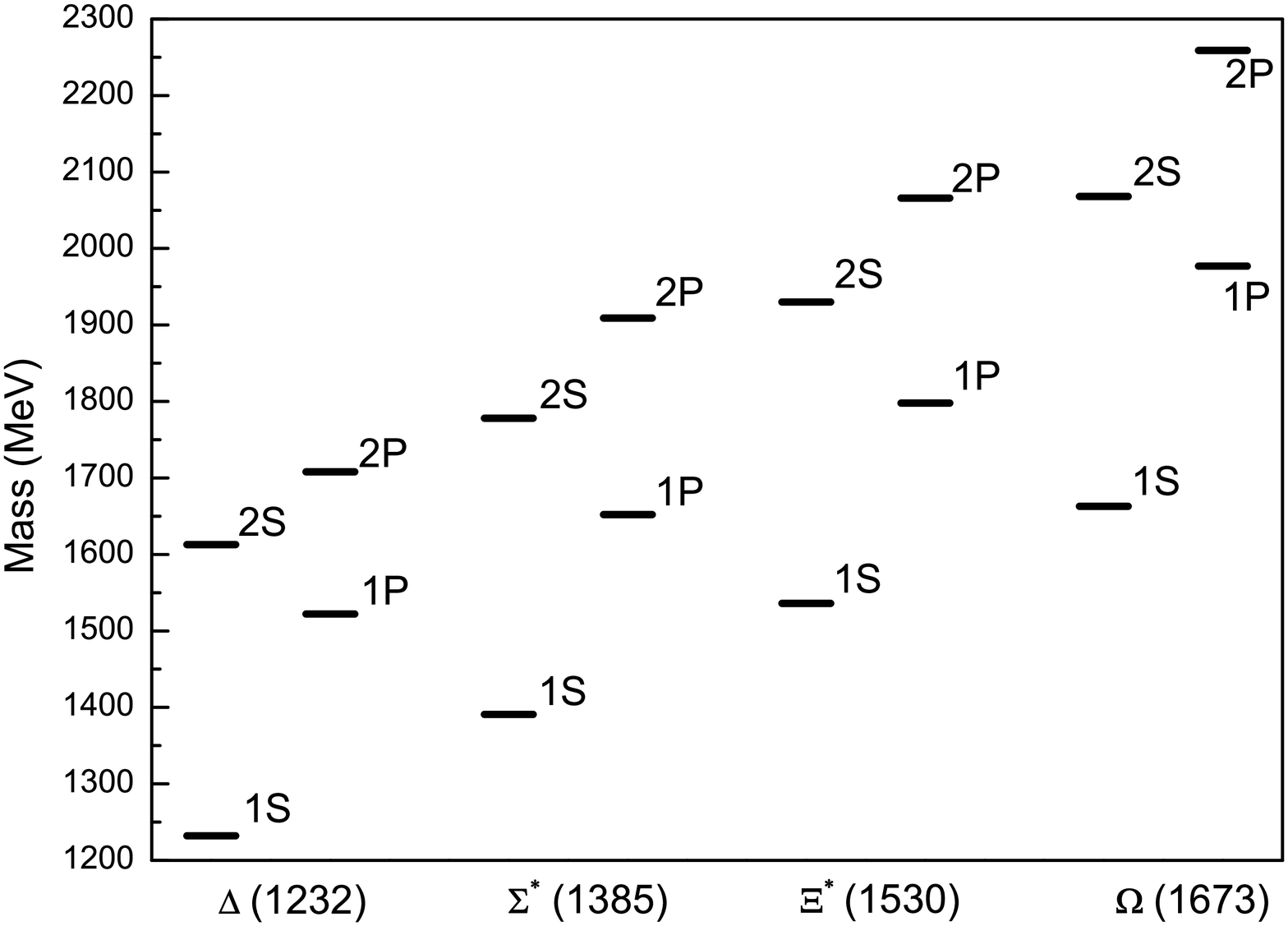} \\[-4ex]
\includegraphics[clip,width=0.49\textwidth,height=0.30\textheight]{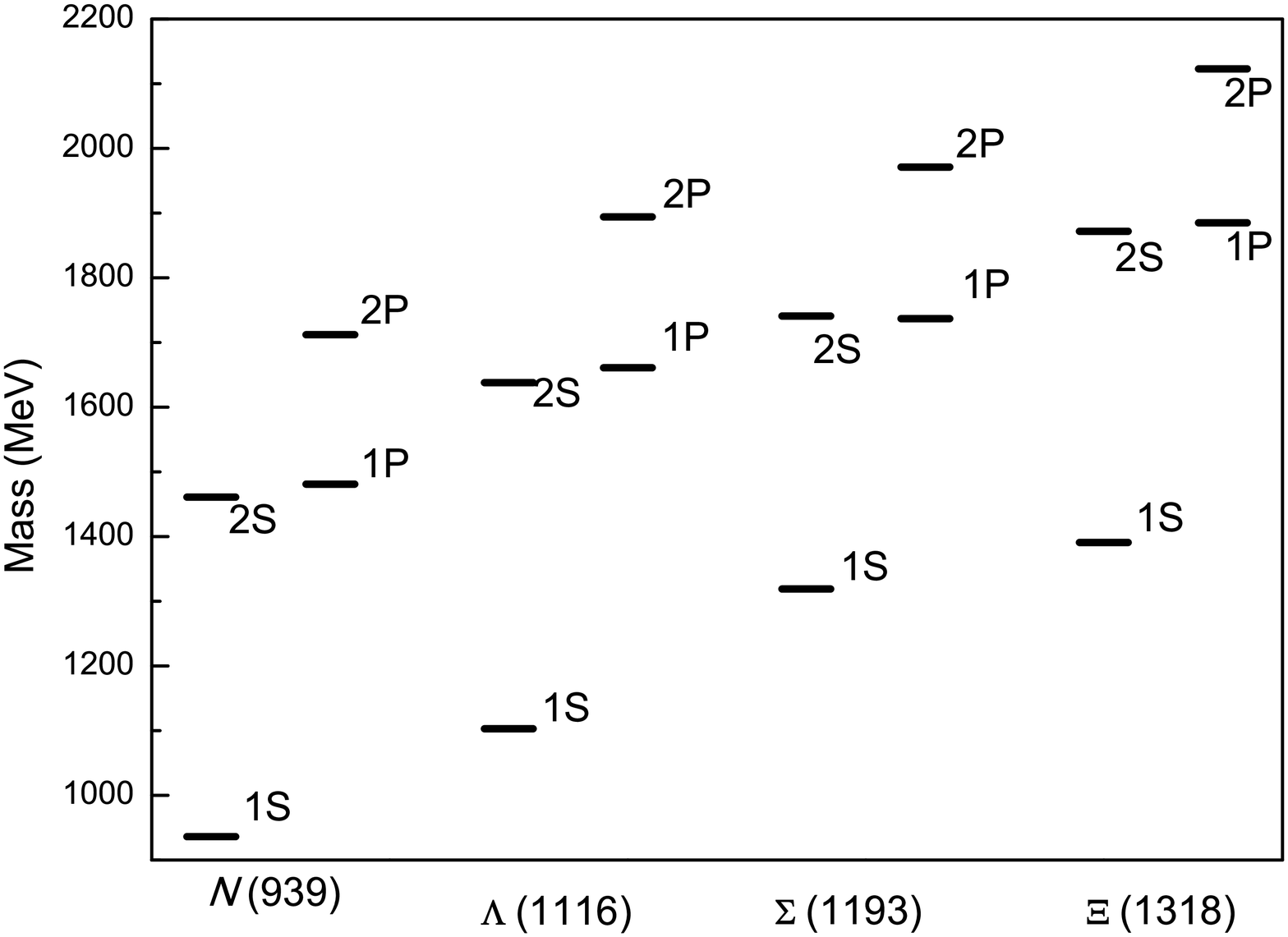}
\put(-130,130){(c)}
\put(35,130){(d)}
\includegraphics[clip,width=0.49\textwidth,height=0.30\textheight]{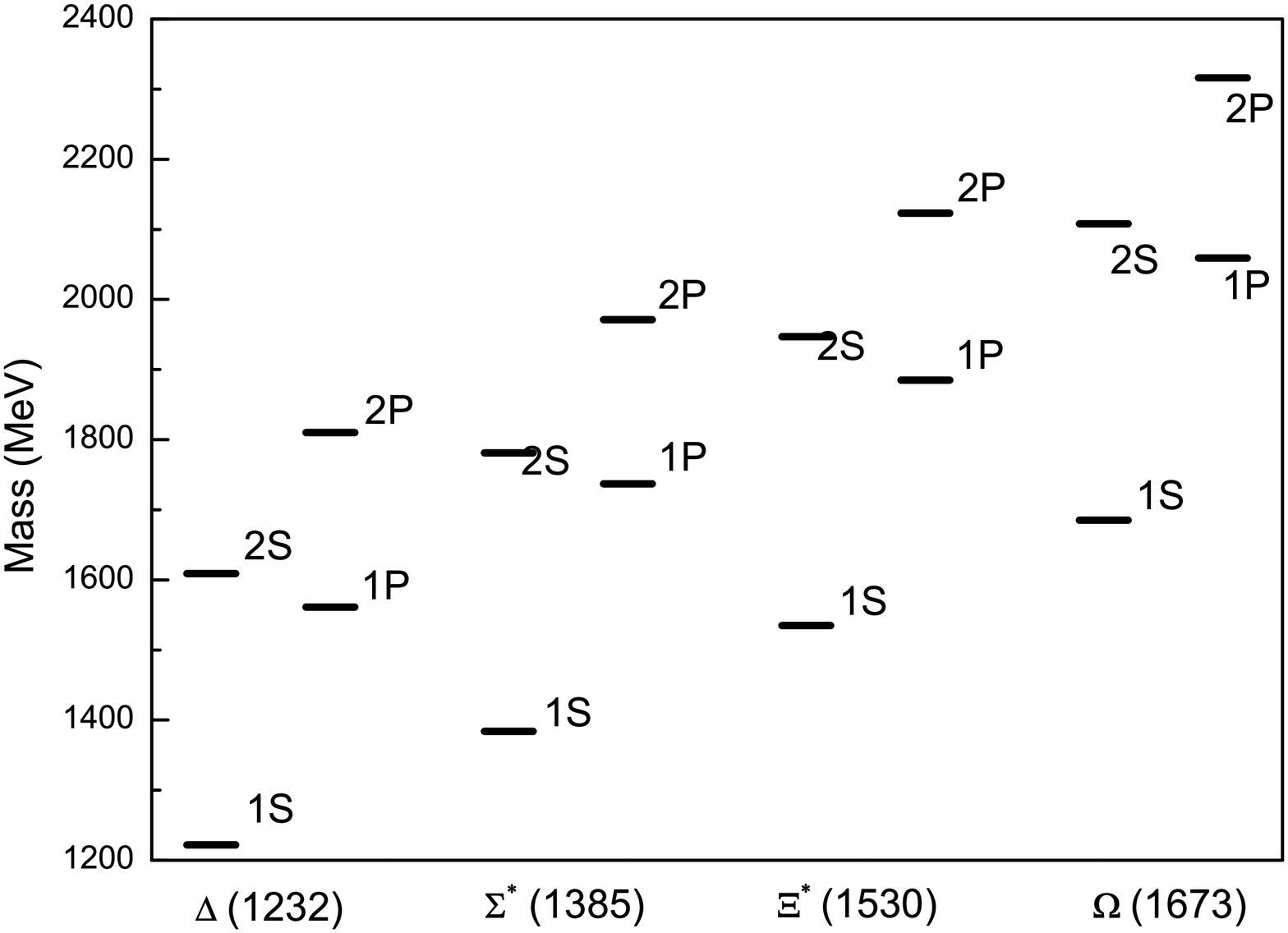}
\caption{\label{fig:spectrum} Masses, in MeV, of the ground and first radial excitation of $S$- and $P$-wave states of the octet and decuplet baryons. {\it Panel (a):} Octet light baryons with set I of model parameters, {\it Panel (b):} Decuplet light baryons with set I of model parameters, {\it Panel (c):} Octet light baryons with set IV of model parameters, and {\it Panel (d):} Octet light baryons with set IV of model parameters.}
\end{center}
\end{figure}

In order to help the reader on guessing the sensitivity of our results with respect the different sets of model parameters, we have drawn in Fig.~\ref{fig:spectrum} the masses presented in Table~\ref{tab:LightBaryons} for the sets I and IV of model parameters. We have opted to plot only the octet and decuplet light baryons because, as will be explained later, the dependence of the mass of heavy baryons with respect the model parameters is milder than in the light quark sector.

\begin{table}[!t]
\begin{center}
\caption{\label{tab:CharmedBaryons} Masses, in MeV, of $1S$, $2S$, $1P$ and $2P$ charmed baryons predicted by the ChQM and using sets I-V of model parameters. Experimental data are taken from Ref.~\cite{Patrignani:2016xqp}.}
\begin{tabular}{ccccccc}
\hline\noalign{\smallskip}
$\Lambda_c(2287)$ & set I & set II & set III & set IV & set V & Exp. \\
$1S$ & 2279 & 2246 & 2291 & 2286 & 2285 & 2287 \\
$2S$ & 2661 & 2654 & 2825 & 2636 & 2657 & 2765 \\
$1P$ & 2547 & 2529 & 2632 & 2543 & 2568 & 2595 \\
$2P$ & 2798 & 2772 & 2956 & 2764 & 2777 & 2940 \\[1.5ex]
$\Sigma_{c}(2454)$ & set I & set II & set III & set IV & set V & Exp. \\
$1S$ & 2454 & 2413 & 2467 & 2434 & 2428 & 2454 \\
$2S$ & 2806 & 2769 & 2976 & 2757 & 2761 & 2800 \\
$1P$ & 2706 & 2665 & 2793 & 2669 & 2681 & 2765 \\
$2P$ & 2960 & 2911 & 3190 & 2894 & 2880 & - \\[1.5ex]
$\Xi_{c}(2470)$ & set I & set II & set III & set IV & set V & Exp. \\
$1S$ & 2522 & 2508 & 2521 & 2569 & 2559 & 2470 \\
$2S$ & 2888 & 2874 & 3025 & 2895 & 2911 & 2930 \\
$1P$ & 2782 & 2778 & 2842 & 2811 & 2829 & 2790 \\
$2P$ & 3042 & 3019 & 3236 & 3034 & 3042 & 3055 \\[1.5ex]
$\Xi_{c}^\prime(2578)$ & set I & set II & set III & set IV & set V & Exp. \\
$1S$ & 2587 & 2571 & 2578 & 2599 & 2580 & 2578 \\
$2S$ & 2943 & 2925 & 3073 & 2920 & 2925 & 2930 \\
$1P$ & 2846 & 2835 & 2895 & 2835 & 2841 & 2790 \\
$2P$ & 3100 & 3074 & 3284 & 3057 & 3052 & 3080 \\[1.5ex]
$\Omega_{c}(2695)$ & set I & set II & set III & set IV & set V & Exp. \\
$1S$ & 2704 & 2705 & 2693 & 2758 & 2724 & 2695 \\
$2S$ & 3070 & 3078 & 3184 & 3083 & 3084 & - \\
$1P$ & 2979 & 2978 & 3010 & 3001 & 2997 & - \\
$2P$ & 3237 & 3233 & 3396 & 3222 & 3218 & - \\[1.5ex]
$\Sigma_{c}^\ast(2520)$ & set I & set II & set III & set IV & set V & Exp. \\
$1S$ & 2486 & 2459 & 2481 & 2449 & 2436 & 2520 \\
$2S$ & 2824 & 2788 & 2984 & 2763 & 2764 & 2800 \\
$1P$ & 2706 & 2665 & 2793 & 2669 & 2681 & 2765 \\
$2P$ & 2960 & 2911 & 3190 & 2894 & 2880 & - \\[1.5ex]
$\Xi_{c}^\ast(2645)$ & set I & set II & set III & set IV & set V & Exp. \\
$1S$ & 2625 & 2613 & 2593 & 2615 & 2589 & 2645 \\
$2S$ & 2964 & 2946 & 3082 & 2927 & 2929 & 2980 \\
$1P$ & 2782 & 2778 & 2842 & 2811 & 2829 & 2790 \\
$2P$ & 3042 & 3019 & 3236 & 3034 & 3042 & 3055 \\[1.5ex]
$\Omega_{c}^\ast(2770)$ & set I & set II & set III & set IV & set V & Exp. \\
$1S$ & 2747 & 2759 & 2708 & 2777 & 2736 & 2770 \\
$2S$ & 3096 & 3101 & 3194 & 3091 & 3089 & - \\
$1P$ & 2979 & 2978 & 3010 & 3001 & 2997 & - \\
$2P$ & 3237 & 3233 & 3396 & 3222 & 3218 & - \\
\noalign{\smallskip}\hline
\end{tabular}
\end{center}
\end{table}

\begin{table}[!t]
\begin{center}
\caption{\label{tab:2CharmedBaryons} Masses, in MeV, of $1S$, $2S$, $1P$ and $2P$ doubly-charmed baryons predicted by the ChQM and using sets I-V of model parameters. Experimental data are taken from Ref.~\cite{Patrignani:2016xqp}.}
\begin{tabular}{ccccccc}
\hline\noalign{\smallskip}
$\Xi_{cc}(3519)$ & set I & set II & set III & set IV & set V & Exp. \\
$1S$ & 3580 & 3572 & 3540 & 3589 & 3544 & 3621 \\
$2S$ & 3906 & 3880 & 3957 & 3849 & 3836 & - \\
$1P$ & 3827 & 3812 & 3811 & 3779 & 3759 & - \\
$2P$ & 4060 & 4024 & 4146 & 3969 & 3956 & - \\[1.5ex]
$\Omega_{cc}$ & set I & set II & set III & set IV & set V & Exp. \\
$1S$ & 3692 & 3721 & 3638 & 3758 & 3696 & - \\
$2S$ & 4042 & 4049 & 4075 & 4033 & 4011 & - \\
$1P$ & 3957 & 3979 & 3921 & 3963 & 3930 & - \\
$2P$ & 4214 & 4203 & 4274 & 4162 & 4142 & - \\[1.5ex]
$\Xi_{cc}^\ast$ & set I & set II & set III & set IV & set V & Exp. \\
$1S$ & 3625 & 3625 & 3556 & 3604 & 3552 & - \\
$2S$ & 3930 & 3913 & 3967 & 3859 & 3842 & - \\
$1P$ & 3827 & 3812 & 3811 & 3779 & 3759 & - \\
$2P$ & 4060 & 4024 & 4146 & 3969 & 3956 & - \\[1.5ex]
$\Omega_{cc}^\ast$ & set I & set II & set III & set IV & set V & Exp. \\
$1S$ & 3742 & 3777 & 3654 & 3777 & 3708 & - \\
$2S$ & 4072 & 4084 & 4085 & 4045 & 4018 & - \\
$1P$ & 3957 & 3979 & 3921 & 3963 & 3930 & - \\
$2P$ & 4214 & 4203 & 4274 & 4162 & 4142 & - \\
\noalign{\smallskip}\hline
\end{tabular}
\end{center}
\end{table}

\begin{table}[!t]
\begin{center}
\caption{\label{tab:BottomBaryons} Masses, in MeV, of $1S$, $2S$, $1P$ and $2P$ single-bottom baryons predicted by the ChQM and using sets I-V of model parameters. Experimental data are taken from Ref.~\cite{Patrignani:2016xqp}.}
\begin{tabular}{ccccccc}
\hline\noalign{\smallskip}
$\Lambda_b(5620)$ & set I & set II & set III & set IV & set V & Exp. \\
$1S$ & 5617 & 5616 & 5622 & 5639 & 5652 & 5620 \\
$2S$ & 5987 & 5996 & 6127 & 5982 & 6023 & - \\
$1P$ & 5875 & 5893 & 5941 & 5891 & 5931 & - \\
$2P$ & 6137 & 6140 & 6298 & 6120 & 6153 & - \\[1.5ex]
$\Sigma_{b}(5810)$ & set I & set II & set III & set IV & set V & Exp. \\
$1S$ & 5812 & 5812 & 5816 & 5799 & 5804 & 5810 \\
$2S$ & 6145 & 6148 & 6285 & 6109 & 6134 & - \\
$1P$ & 6039 & 6040 & 6104 & 6020 & 6048 & - \\
$2P$ & 6286 & 6283 & 6480 & 6240 & 6255 & - \\[1.5ex]
$\Xi_{b}(5795)$ & set I & set II & set III & set IV & set V & Exp. \\
$1S$ & 5850 & 5875 & 5843 & 5922 & 5924 & 5795 \\
$2S$ & 6203 & 6232 & 6315 & 6237 & 6271 & - \\
$1P$ & 6099 & 6135 & 6141 & 6155 & 6185 & - \\
$2P$ & 6351 & 6374 & 6507 & 6370 & 6403 & - \\[1.5ex]
$\Xi_{b}^\prime(5935)$ & set I & set II & set III & set IV & set V & Exp. \\
$1S$ & 5937 & 5962 & 5924 & 5962 & 5952 & 5935 \\
$2S$ & 6270 & 6296 & 6370 & 6266 & 6288 & - \\
$1P$ & 6168 & 6204 & 6198 & 6182 & 6200 &  - \\
$2P$ & 6413 & 6433 & 6558 & 6394 & 6415 & - \\[1.5ex]
$\Omega_{b}(6046)$ & set I & set II & set III & set IV & set V & Exp. \\
$1S$ & 6045 & 6102 & 6038 & 6123 & 6096 & 6046\\
$2S$ & 6387 & 6443 & 6471 & 6425 & 6440 & - \\
$1P$ & 6288 & 6339 & 6303 & 6343 & 6348 & - \\
$2P$ & 6536 & 6582 & 6656 & 6552 & 6571 & - \\[1.5ex]
$\Sigma_{b}^\ast(5830)$ & set I & set II & set III & set IV & set V & Exp. \\
$1S$ & 5824 & 5829 & 5818 & 5805 & 5807 & 5830 \\
$2S$ & 6151 & 6155 & 6286 & 6112 & 6135 & - \\
$1P$ & 6039 & 6040 & 6104 & 6020 & 6048 & - \\
$2P$ & 6286 & 6283 & 6480 & 6240 & 6255 & - \\[1.5ex]
$\Xi_{b}^\ast(5955)$ & set I & set II & set III & set IV & set V & Exp. \\
$1S$ & 5952 & 5978 & 5927 & 5969 & 5956 & 5955 \\
$2S$ & 6278 & 6304 & 6371 &6269 & 6290 & - \\
$1P$ & 6099 & 6135 & 6141 & 6155 & 6185  & - \\
$2P$ & 6351 & 6374 & 6507 & 6370 & 6403 & - \\[1.5ex]
$\Omega_{b}^\ast$ & set I & set II & set III & set IV & set V & Exp. \\
$1S$ & 6062 & 6121 & 6040 & 6131 & 6101 & - \\
$2S$ & 6397 & 6451 & 6473 & 6429 & 6442 & - \\
$1P$ & 6288 & 6339 & 6303 & 6343 & 6348 & - \\
$2P$ & 6536 & 6582 & 6656 & 6552 & 6571 & - \\
\noalign{\smallskip}\hline
\end{tabular}
\end{center}
\end{table}

\begin{table}[!t]
\begin{center}
\caption{\label{tab:2BottomBaryons} Masses, in MeV, of $1S$, $2S$, $1P$ and $2P$ double-bottom baryons predicted by the ChQM and using sets I-V of model parameters. Experimental data are taken from Ref.~\cite{Patrignani:2016xqp}.}
\begin{tabular}{ccccccc}
\hline\noalign{\smallskip}
$\Xi_{bb}$ & set I & set II & set III & set IV & set V & Exp. \\
$1S$ & 10084 & 10314 & 10064 & 10308 & 10271 & - \\
$2S$ & 10412 & 10542 & 10374 & 10492 & 10494 & - \\
$1P$ & 10358 & 10483 & 10271 & 10439 & 10427 & - \\
$2P$ & 10555 & 10651 & 10518 & 10579 & 10592 & - \\[1.5ex]
$\Omega_{bb}$ & set I & set II & set III & set IV & set V & Exp. \\
$1S$ & 10179 & 10462 & 10161 & 10481 & 10424 & - \\
$2S$ & 10525 & 10709 & 10483 & 10680 & 10666 & - \\
$1P$ & 10470 & 10644 & 10375 & 10623 & 10593 & - \\
$2P$ & 10683 & 10824 & 10634 & 10757 & 10740 & - \\[1.5ex]
$\Xi_{bb}^\ast$ & set I & set II & set III & set IV & set V & Exp. \\
$1S$ & 10123 & 10336 & 10068 & 10314 & 10275 & - \\
$2S$ & 10424 & 10558 & 10376 & 10497 & 10497 & - \\
$1P$ & 10358 & 10483 & 10271 & 10439 & 10427 & - \\
$2P$ & 10555 & 10651 & 10518 & 10579 & 10592 & - \\[1.5ex]
$\Omega_{bb}^\ast$ & set I & set II & set III & set IV & set V & Exp. \\
$1S$ & 10216 & 10486 & 10164 & 10490 & 10429 & - \\
$2S$ & 10540 & 10724 & 10485 & 10686 & 10669 & - \\
$1P$ & 10470 & 10644 & 10375 & 10623 & 10593 & - \\
$2P$ & 10683 & 10824 & 10634 & 10757 & 10740 & - \\
\noalign{\smallskip}\hline
\end{tabular}
\end{center}
\end{table}

Dynamical chiral symmetry breaking effects are much less important, even negligible, when heavy quarks are present. This translates in our formalism to the fact that the interaction terms between light-light, light-heavy and heavy-heavy quarks are not the same. For example, while Goldstone-boson exchanges are considered when the two quarks are light, they do not appear in the other two configurations: light-heavy and heavy-heavy. This must have consequences in the way the $2S$ and $1P$ states are located in the spectrum of heavy baryons and thus we consider this study relevant for our discussion. Furthermore, many states in the $qqQ$ and $qQQ$ ($q$ represents a light $u$-, $d$-, or $s$-quark and $Q$ denotes either $c$- or $b$-quark) heavy baryon sectors have been recently reported, forcing us to present our predictions.

Either charmed or bottom baryons have been studied theoretically during the last few decades. After the discovery of the first charmed baryons, several phenomenological potential models, developed for describing the light baryon and/or meson spectra, were applied to analyze the properties of observed and expected heavy baryon states~\cite{Copley:1979wj, Maltman:1980er, Richard:1983tc, Capstick:1986bm, SilvestreBrac:1996bg, Ebert:2005xj, Ebert:2007nw, Roberts:2007ni, Valcarce:2008dr}. Beyond the quark model approach, Roncaglia et al.~\cite{Roncaglia:1995az} predicted the masses of baryons containing one or two heavy quarks using the Feynman-Hellmann theorem and semi-empirical mass formulae. Jenkins~\cite{Jenkins:1996de} studied heavy baryon masses in a combined expansion in $1/m_{Q}$, $1/N_{c}$, and $SU(3)$ flavor symmetry breaking. The QCD sum rules approach has been applied to the spectra of heavy baryons in Refs.~\cite{Bagan:1991sc, Bagan:1992tp, Wang:2002ts, Wang:2007sqa, Liu:2007fg}. A preliminary description of heavy baryons based on the Dyson-Schwinger Equations formalism has been released~\cite{Qin:2018dqp}. Within lattice-regularised QCD techniques, Bowler et al.~\cite{Bowler:1996ws} made an exploratory study of charmed and bottom baryons. Mathur et al.~\cite{Mathur:2002ce} gave a more precise prediction of their masses using the quenched approximation. And, finally, Brown et al.~\cite{Brown:2014ena} published one of the most up-to-date lattice QCD calculations of masses of baryons containing one, two, or three heavy quarks in any possible combination.

No spin or parity quantum numbers of a heavy baryon candidate have been measured experimentally, but they are assigned based on quark model expectations. Such properties can only be extracted by studying angular distributions of the particle decays, that are available only for the lightest and most abundant species. For excited heavy baryons the data set is typically one order of magnitude smaller than for heavy mesons and therefore the knowledge of radially and orbitally excited states is very much limited. All together explains why it is nowadays difficult to check the order of the $2S$ and $1P$ states. In our formalism, as one can see in the second column of Tables~\ref{tab:CharmedBaryons},~\ref{tab:2CharmedBaryons},~\ref{tab:BottomBaryons} and~\ref{tab:2BottomBaryons}, the radial excitation is located above the orbital-excited state in all the spin-parity channels considered. This situation is also found for the remaining four sets of model parameters and it is in agreement with the results of, for instance, Refs.~\cite{Capstick:1986bm, Valcarce:2008dr}. Our theoretical results for heavy baryons are more stable when going from set I to set V of model parameters; the only exception is the $\Xi_b$ baryon whose mass varies from $5.84$ GeV to $5.92$ GeV. It is worth noting that there is no contribution from the Goldstone-boson exchange interactions to the doubly--heavy-flavor hadrons, hence it is quite different from that of nucleon or $\Lambda$ baryon where the chiral potential plays an important role for the mass ordering.

We proceed now to discuss possible assignments of our theoretical states to the ones already observed experimentally. Within the charmed baryon sector, as one can observe in Tables~\ref{tab:CharmedBaryons} and~\ref{tab:2CharmedBaryons}, the theoretical masses of all ground states are in reasonably good agreement with the ones collected in PDG~\cite{Patrignani:2016xqp}. Let us highlight our result for the recently discovered $\Xi_{cc}^{++}$ baryon with a theoretical mass of $3589\,{\rm MeV}$ in set IV, in reasonable agreement with its experimental measurement: $(3621\pm0.77)\,\text{MeV}$~\cite{Aaij:2017ueg}. This gives us confidence on the location of the subsequent ground states of the $\Omega_{cc}$, $\Xi_{cc}^{\ast}$ and $\Omega_{cc}^{\ast}$ baryons. Our predictions are in the energy ranges $3.64-3.75$ GeV, $3.55-3.62$ GeV and $3.65-3.77$ GeV, respectively.

The assignment of excited states to the remaining charmed baryons collected in the PDG is avoided herein because, as mentioned above, the experimental situation is complex. Note, however, that the LHCb Collaboration has recently announced five new excited $\Omega_c^0$ states: $\Omega_c(3000)^0$, $\Omega_c(3050)^0$, $\Omega_c(3066)^0$, $\Omega_c(3090)^0$ and $\Omega_c(3119)^0$~\cite{Aaij:2017nav}. In our chiral quark model, and using all sets of model parameters, both $2S$ and $1P$ states of the $\Omega_{c}$ and $\Omega_{c}^{\ast}$ baryons are predicted to be in the same energy region and thus some of them appear as natural candidates; this conclusion also had been proposed in Ref.~\cite{Yang:2017rpg}.

The experimental data is more scarce for the bottom baryons than for the charmed ones: not all the ground states are well established and only three excited states have been detected until now~\cite{Patrignani:2016xqp}. We predict a ground state $\Lambda_b$-baryon located at the energy region $5.62-5.65$ GeV and the one of $\Sigma_b$ at $5.80-5.81$ GeV, these values are compatible with experiment ($\Lambda_b(5620)$ and $\Sigma_b(5810)$), and a similar situation is found for the $\Omega_b$: $6.04-6.12$ GeV whose experimental data is $6046$ MeV. In contrast, the energy interval of $5.84-5.92$ GeV is slightly higher than the experimental mass of the ground state $\Xi_b$-baryon. After looking over the five sets of results, the predicted intervals for the ground states of the bottom baryons $\Sigma_b^\ast(5830)$, $\Xi_b^\prime(5935)$ and $\Xi_b^\ast(5955)$ are $5.81-5.82$ GeV, $5.93-5.96$ GeV and $5.93-5.98$ GeV, respectively. All of them are in reasonably good agreement with the masses collected in the PDG.

Doubly bottom baryons have not yet been observed. Following the experimental situation in the corresponding charmed sector, we expect that the ground state of the $\Xi_{bb}$ baryon will be firstly observed. We predict a mass for such state between $10.06$ GeV and $10.31$ GeV; note that Ref.~\cite{Valcarce:2008dr} predicts a mass for the $\Xi_{bb}$ ground state which is located within our interval. Attending to columns 2-5 of Table~\ref{tab:2BottomBaryons}, the ground state mass for the $\Omega_{bb}$, $\Xi_{bb}^\ast$ and $\Omega_{bb}^\ast$ baryons are located, respectively, within the intervals $10.16-10.48$ GeV, $10.07-10.33$ GeV and $10.16-10.49$ GeV.

Once the light, one-heavy and two-heavy baryons have been computed within a common framework, it is interesting to look for patterns that relate different baryon sectors. As one can see from Table~\ref{tab:LightBaryons} to~\ref{tab:2BottomBaryons}, the first pattern that we find is that the mass difference between octet and decuplet baryons is smaller when heavy quarks are involved. Another interesting feature is that the $1S$-$2S$ and $1S$-$1P$ mass splittings tend to be smaller when the mass of the heavy quarks is larger. This kind of patterns are related with the heavy quark mass expansion~\cite{Jenkins:1996de}.


\section{Summary}
\label{sec:summary}

Masses of the $1S$, $2S$, $1P$ and $2P$ states of light, charmed, doubly-charmed, single-bottom and double-bottom baryons have been presented, paying particular attention to the $1P$-$2S$ splittings in order to guess possible solutions of the so-called level ordering problem in the nucleon spectrum.

We have used a chiral quark model that contains, besides the perturbative one-gluon exchange interaction and a linear-screened confining potential, Goldstone-boson exchange potentials between quarks. In the meson-exchange interactions, we have considered the full octet of pseudoscalar and scalar mesons. The later ones simulate the multi-pion exchange terms that appear in the chiral Lagrangian. They were incorporated in the original model for describing, e.g., the $\rho-\omega$ splitting but not yet to address the level ordering problem. The three-body bound state problem has been solved by means of the Gau\ss ian expansion method which is as accurate as Faddeev calculations. The Gau\ss ian ranges are set in geometric progression, this enables the optimization of them employing a small number of free parameters. As it is well know, the quark model parameters are crucial in order to describe particular physical observables. We have used values that have been fitted before through hadron and hadron-hadron phenomenology. Moreover, we have provided results with five different sets of model parameters in order to get some insight about the uncertainties associated with the model.

We have found that the ground states of the octet and decuplet light baryons are reasonably well described. The pseudoscalar, confining and color Fermi-Breit interactions compete for mass-splittings in the light baryon spectrum and thus, with particular, but still natural, sets of model parameters one could reproduce the $1P$-$2S$ splitting observed experimentally. No one of the mentioned interactions is able to reproduce the $1P$-$2S$ mass-splitting when it is considered alone. The incorporation of the so-called $\sigma$-, $f_{0}$-, $a_{0}$- and $\kappa$-exchange potentials helps on getting a better global description of the light baryon spectrum. It is worth also to note that the screened-linear potential, which mimics the effect of meson-baryon thresholds that are far from the bare (undressed) quark-model states, tends to reduce the mass splitting between the $1S$ and $2S$ states.

Meson-exchange potentials are not considered when the two quarks inside a baryon are either light-heavy or heavy-heavy. This makes the single- and doubly-heavy baryons particularly interesting for our study on the mass-splitting between radially-excited and orbitally-excited states. Furthermore, many states in the $qqQ$ and $qQQ$ ($q$ represents a light $u$-, $d$-, or $s$-quark and $Q$ denotes either $c$- or $b$-quark) heavy baryon sectors have been recently reported. We have observed that the theoretical masses of all ground states are in reasonably good agreement with the ones collected in PDG. The assignment of excited states to PDG candidates has been avoided because the experimental situation is complex. However, some important remarks can be done: (i) the $2S$ state lies above the $1P$ state in all studied spin-parity channels and within any heavy baryon sector; (ii) both the $2S$ and the $1P$ states of the $\Omega_{c}$ and $\Omega_{c}^{\ast}$ baryons are predicted to be in the same energy region than the five $\Omega_c^0$ states recently discovered by the LHCb Collaboration; (iii) the predicted mass of ground state of the $\Xi_{cc}$ baryon is comparable with its recent experimental data; and (iv) the mass splittings appear to be smaller as soon as more valence heavy quarks are present in the description of a baryon.


\vspace*{-0.20cm}
\begin{acknowledgements}
We are grateful for constructive comments from Craig D. Roberts, A. Valcarce, J. Vijande, and F. Wang. Work supported by: National Natural Science Foundation of China under Grant nos. 11535005 and 11775118; European Union's Horizon 2020 research and innovation programme under the Marie Sk\l{}odowska-Curie grant agreement no. 665919; Spanish MINECO's Juan de la Cierva-Incorporaci\'on programme with grant agreement no. IJCI-2016-30028; and by Spanish Ministerio de Econom\'ia, Industria y Competitividad under contract nos. FPA2014-55613-P, FPA2017-86989-P and SEV-2016-0588.
\end{acknowledgements}

\vspace*{-0.50cm}


\bibliographystyle{apsrev}       

%
%
%

\end{document}